# Optical wavefront shaping in deep tissue using photoacoustic feedback

**Fei Xia[1,*], Ivo Leite[2,*], Robert Prevedel[2], and Thomas Chaigne[3]**

[1] Laboratoire Kastler Brossel, ENS–Universite PSL, CNRS, Sorbonne Université, Collège de France, 24 Rue Lhomond, F-75005 Paris, France
[2] European Molecular Biology Laboratory. Meyerhofstraße 1, 69117 Heidelberg, Germany
[3] Aix Marseille Univ, CNRS, Centrale Med, Institut Fresnel, Marseille, France
* These authors contributed equally.

E-mail: thomas.chaigne@fresnel.fr

## Abstract

Over the past decade, optical wavefront shaping has been developed to focus light through highly opaque scattering layers, opening new possibilities for biomedical applications. To probe light intensity deep *inside* soft scattering media such as biological tissues, internal guide-stars are required. Here, we give an overview of the main principles and describe in depth the use of a photoacoustic feedback signal for this purpose. We further present first principles calculations and simulations to estimate important experimental parameters, and detailed instructions on designing and conducting these experiments. Finally, we provide guidance towards selecting suitable equipment for building a typical experimental setup, paving the way for further innovative biomedical imaging and therapy applications.

## 1. Context

In recent years, the field of optical imaging has witnessed significant advancements, particularly in navigating and imaging through highly scattering media such as thick biological tissue [1,2]. A pivotal breakthrough in this arena is the introduction of optical wavefront shaping, a technique that emerged in the last decades, which enables the highly controlled focusing of coherent light through scattering materials. This is accomplished by compensating for the phase distortions that coherent light beams accumulate as they propagate through such media. To determine and apply this phase correction, two main elements are required: 1) a detector providing a feedback signal related to the light intensity at the desired focus location and 2) a spatial light modulator (SLM) to tune the optical wavefront. Spatial light modulators include deformable mirrors, liquid crystal arrays, and digital micromirror devices. They can be used to shape the phase and-or the amplitude of the optical wavefront impinging on the scattering layer. The appropriate correction patterns are usually derived by optimizing the feedback signal and therefore the light intensity [3]. Alternatively, the so-called transmission (or reflection) matrix can also be measured to extract these phase corrections [4].

While focusing *through* a scattering medium can be somewhat straightforward—owing to the ability to place detectors on the medium's opposite side to capture feedback—focusing *within* a spatially extended medium itself poses more challenges. It necessitates a proxy to measure light intensity at depth, ushering in the utilization of so-called guide-star mechanisms [5]. For instance, correction phase patterns maximising the fluorescence [6,7] or SHG [8] emission of nanoparticles embedded inside a scattering medium enabled to concentrate light on these emitters. However this requires to insert these probes inside the scattering medium, and light can then only be focused in their vicinity. Additionally, penetration capability is limited as sufficient feedback signal must be generated and detected to efficiently start the optimization procedure.

Photoacoustic (PA) (also know as optoacoustic) feedback, has also emerged about a decade ago as a promising guide-star mechanism for wavefront shaping [9,10]. This approach leverages the generation of acoustic waves from the rapid thermal expansion caused by transient light absorbed by a material [11]. Notably, unlike visible light, these acoustic waves can propagate through soft scattering tissue with minimal scattering, and can be measured in a time-resolved manner. This provides a significant advantage over traditional optical signal measurements, which are generally limited to intensity and restricted in depth due to absorption and-or scattering. The technique not only encodes information about the optical properties of materials, such as biological tissues, but also combines this high-specificity of optical imaging with the deep penetration capabilities of ultrasound. This blend is invaluable in fields such as biomedical imaging. There, the trade-off between depth and resolution (inversely proportional to the ultrasound frequency) in PA imaging is fundamentally governed by the attenuation of ultrasound in tissues, which typically increases by approximately 1 dB per cm per MHz. It is also constrained by the maximum level of optical energy that biological tissue can tolerate without sustaining damage. This yields a typical depth-resolution ratio of about 200: a resolution of approximately 100 micrometers (for detected ultrasound frequencies around 15 MHz) can be obtained at depths up to 2 centimeters within biological tissues [12].

The use of PA feedback in wavefront shaping involves the use of diffuse coherent light fields, or speckle patterns, that upon encountering an optically absorbing object within soft tissue, result in a modulation of the PA signal. This modulation serves as a crucial feedback signal for implementing optical wavefront shaping methods, aiming to focus light inside scattering media. The potential of this technique has been explored over the past decade, and light focusing through scattering media has been demonstrated using single-element ultrasonic transducer or multidimensional arrays.

Although the resolution of the optical focus achieved is generally dictated by ultrasound resolution [13], the broadband nature of the photoacoustic signal [14], the sensitivity profile of focused sensors [15] and thermo-acoustic non-linearities [16] have been leveraged to further increase the focusing capabilities of this technique, down to the optical diffraction limit [16].

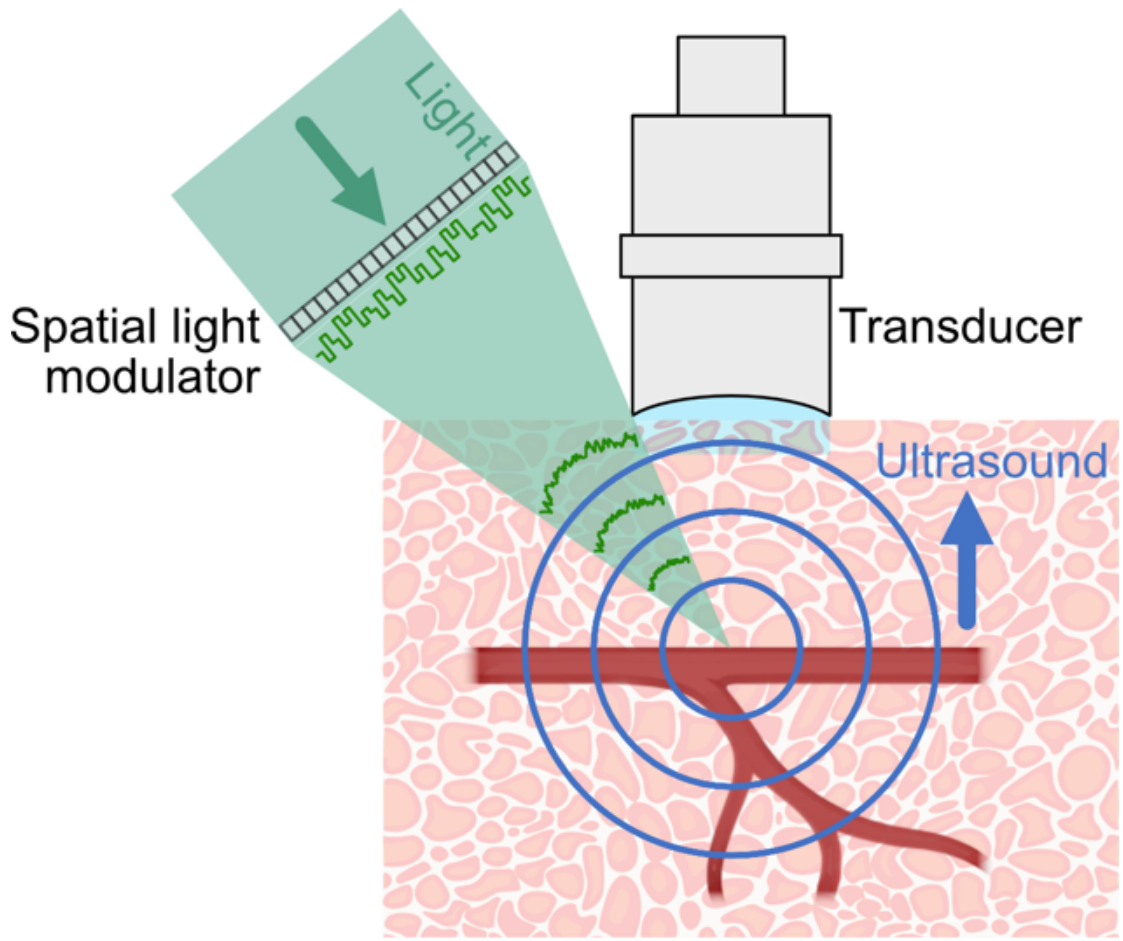

**Figure 1 - Principle of PA-guided optical wavefront shaping**. The wavefront of a coherent light source is shaped with a spatial light modulator to pre-compensate the distortions when propagating in scattering media (such as biological tissue). The applied phase pattern is optimized to maximize the amplitude of the ultrasound generated by optically absorbing structures upon pulsed illumination.

This tutorial is part of the " Focus Issue on Foundational Skills and Tools for Building Wavefront Shaping Systems " series. Here we aim to specifically demystify the concepts and methodologies underpinning PA-guided wavefront shaping. By elucidating the principles, methods and tools, challenges, and potential applications, we seek to provide a comprehensive overview that can serve as a foundational resource for researchers and practitioners in the field. We provide clear, step-by-step instructions for setting up and conducting experiments using PA feedback. The guide covers selecting the right equipment, preparing samples, along with processing and interpreting PA signals. We also provide some simulation tools to test various experimental configurations and focusing strategies. As a resource for newcomers to PA-based wavefront shaping, this paper offers a straightforward approach to understanding and applying this advanced light manipulation technique. Finally, we discuss the challenges to focus light *inside soft* scattering media and potential solutions.

## 2. Experimental system design and procedure

### *2.1 Experimental setup*

In Figure 2.a,b, we show a traditional optical wavefront shaping setup. A light beam is shaped in phase and-or amplitude by a SLM and then propagates through a scattering sample. It generates a speckle pattern due to the interference of different light paths, which can be controlled by shaping the input wavefront with the SLM to optimize a feedback signal. The latter is usually an optical signal at the desired focus location [5].

The experimental configuration of a PA-guided wavefront shaping system shares similarities with it. The primary distinction in a PA-guided wavefront shaping system arises from the use of a PA-induced feedback mechanism (i.e. based on the generation of an acoustic signal through light-induced thermal expansion), which can be efficiently detected using ultrasound sensors such as piezoelectric transducers.

In Figure 2, we describe a standard wavefront shaping setup using a photoacoustic signal as the feedback. Typically, a PA laser source hits the spatial light modulator that modulates the wavefront of the incoming laser beam on the sample. On the detection arm, an ultrasound transducer is immersed in a coupling medium (typically ultrasound gel or water) which picks up the acoustic signal. The detected signal is then used as a feedback signal that guides the wavefront shaping process in the spatial light modulator. In Figures 2.c-e, we show various acoustic detection configurations; depending on the specific application, different designs can be used.

### *2.2 Optimization procedure*

Optimizing the wavefront of the PA-excitation laser is similar to other optical wavefront shaping techniques. We first define optical input modes with a given basis, such as individual SLM pixels, groups of pixels as defined by an Hadamard basis [4], or phase ramps to span the conjugated Fourier basis [17]. The number of optical input modes determines the controllable degree of freedom. Each mode is then modulated (in phase from 0 to $2\pi$, or in amplitude), and the induced modulation in PA feedback signal is detected. The phase/amplitude maximizing this feedback signal is then extracted. By sequentially adjusting the phase or amplitude of these input modes, light is progressively focused. This is typically referred to as an iterative focusing method. Alternatively, transmission matrix approaches can also be implemented with PA feedback signal as an non-iterative method, as further detailed in section 'Which photoacoustic feedback signal?'. As PA feedback detection schemes often suffer from poor SNR and-or faint modulations, it is usually critical to span an input basis for which as many SLM pixels as possible can be modulated. The more degree of freedom are controlled, in principle, the larger the focused light intensity enhancement [3]. In the following section, we describe further the required experimental conditions to successfully perform this optimization procedure.

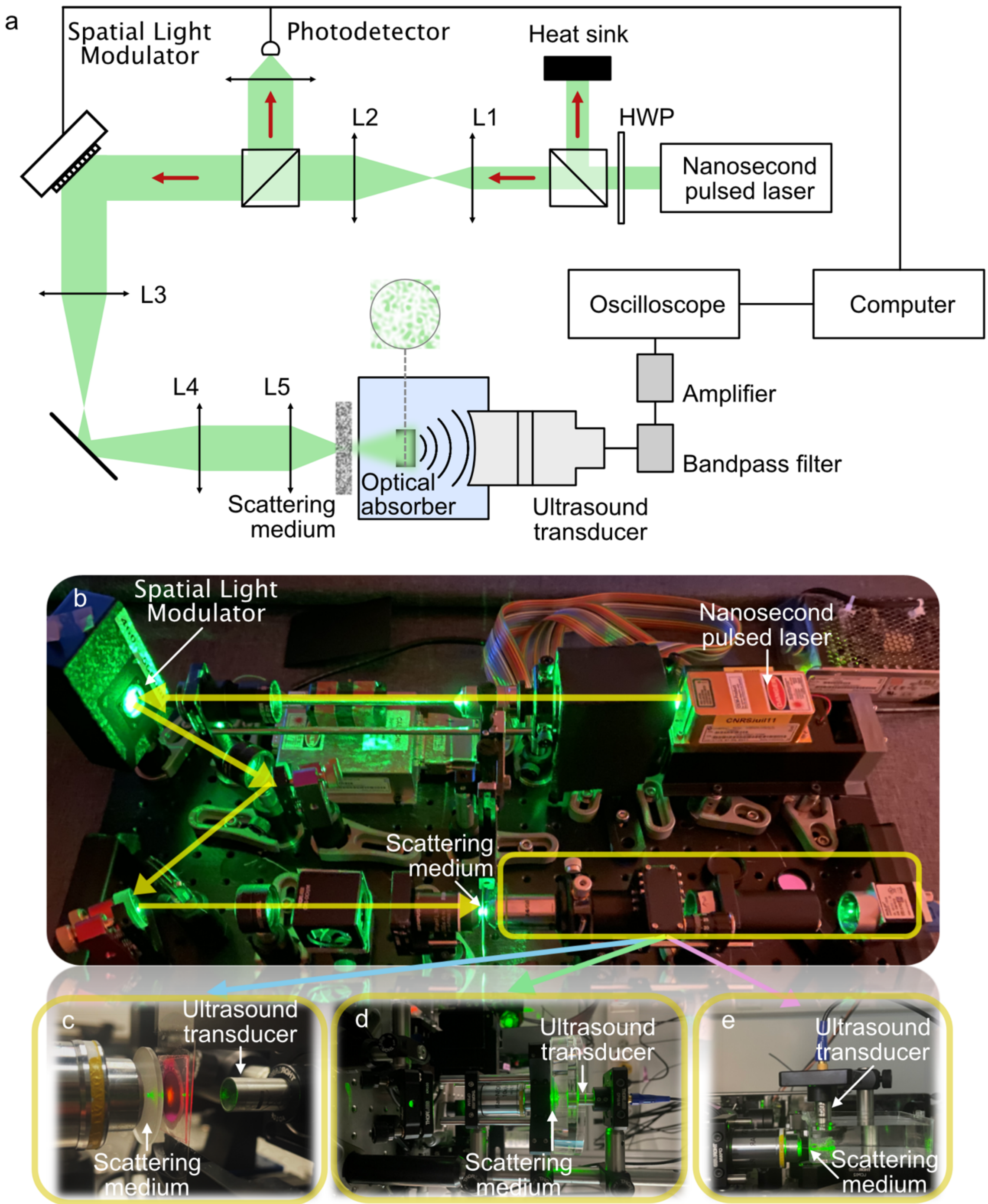

**Figure 2 - Experimental Setup**. **a**, Experimental diagram. HWP: half-wave plate, L1-L5: lenses. **b**, Photo for a generalized wavefront shaping setup involves a laser passing through a telescope beam expander to fill the active area of a spatial light modulator. The light is then reflected and directed into a sample that is either embedded in or placed behind a scattering medium. A detection system, which can vary depending on the signal, is utilized within the boxed area. **c-e**, Various detection configurations for PA signal, the detection arm may include a transducer embedded within ultrasound gel (used for acoustic coupling between the sample and transducer, omitted in the figure for easier visualization). A thin film material, such as FEP or PET foil, can be chosen to serve as a sample cover instead of glass cover slip to minimize ultrasound reflections of PA waves, as it possesses an acoustic impedance similar to water (and ultrasound gel). Alternatively, a water tank with the transducer immersed in the direction of transmission (d) or orthogonal (e) to it can be used. The object generating the acoustic signal often consists of highly absorbing liquid dyes or other absorbing materials.

## 3. Practical experimental considerations

The light focusing performance following the optical wavefront correction is usually assessed as the contrast between the intensity at the focused spot compared to the surrounding speckle grains. For a given number of degrees of freedom (or 'input' modes), there is only a given amount of energy that can be directed towards the focus location. Therefore, the optical intensity enhancement is driven by the number of simultaneously optimized speckle grains, or 'output' modes. In this section, we compute this number to evaluate how many input modes would be required to obtain significant focusing capability *inside* a scattering medium. This highlights the bottlenecks in PA-guided wavefront shaping, as this assessment is critical to predict its performances in realistic scenarios but is often overlooked in the literature.

### *3.1 How many speckle grains are probed?*

Let's consider a targeted focusing depth of 1 mm inside tissue and an illumination wavelength of 1 µm. We assume a fully developed speckle volumetric pattern [18], yielding speckle grains with diffraction limited diameter of approximately $\lambda/2 = 0.5\ \mu m$. As light propagates in a diffusive manner, the radius of the illuminated area in the output plane is roughly given by the thickness of the scattering medium [19]. This means that the illuminated area at a depth of 1 mm is approximately $\pi \times 1$ mm$^2$. In this plane, there are therefore $\frac{\pi \times 1000^2}{\pi \times 0.25^2} = 16 \cdot 10^6$ output modes or speckle grains.

Let's now consider a focused single-element ultrasound transducer, of central frequency $f = 30\ MHz$, with equal focal length $F$ and diameter $D$: $F = D = 10\ mm$. The transverse ($\Delta x$) and axial ($\Delta z$) dimensions of its sensitivity focal region (at -6dB pulse-echo, or -3dB in detection only) are limited by acoustic diffraction, and as such depend on the acoustic wavelength and the aperture of the transducer [20]:

(1) $\Delta x = \frac{c_s}{f} \times \frac{F}{D} = 50\ \mu m$

with the speed of sound in water is $c_s$=1490m/s,

(2) $\Delta z = N_d \times S_F^2 \times \frac{2}{1+0.5S_F} \simeq 400\ \mu m$

with: $N_d = \frac{D^2 f}{4c}$ (near field distance) and $S_F = \frac{F}{N_d}$ (normalized focal length).

Considering a uniformly absorbing transverse plane, the PA signal is effectively probing $N_{speckle} = \frac{\pi(\Delta x/2)^2}{\pi(\lambda/4)^2} = \frac{\Delta x^2}{(\lambda/2)^2} = \frac{50^2}{0.5^2} = 10000$ speckle grains. This sets several constraints on the sensitivity and the signal-to-noise ratio of the PA detection system, as well as on the focusing capability, which will both be discussed later in sections 3.2 and 3.3.

### *3.2 Required signal-to-noise ratio*

The amplitude of the modulation of the PA signal due to the fluctuations must be larger than the noise level. The total light intensity scales with $N_{speckle}$, whereas the modulation amplitude scales with $\sqrt{N_{speckle}}$ [18]. The expected relative modulation amplitude (modulation amplitude divided by peak-to-peak amplitude, see Fig.3c) is therefore:

$$\frac{1}{\sqrt{N_{speckle}}} = \frac{\lambda}{2\Delta x} = 10^{-2}$$

The signal-to-noise ratio (SNR) of the PA detection system must therefore be large enough to detect such a faint modulation. Additionally, one must make sure that the dynamic range of the analog-to-digital converter/digitizer is appropriate to resolve such faint modulation (see also section 4.3).

Several solutions can be considered to reduce the number of probed speckle grains and increase this relative modulation amplitude: first using sparser and smaller absorbers will reduce the effective number of probed speckle grains as well; second increasing the ultrasound detection frequency will reduce the extent of the focal region of the transducer.

Nonetheless, these solutions can exhibit some downside. Sparsity will reduce the total optical absorption, effectively reducing the emitted ultrasound power. Although the emitted PA signals are fundamentally broadband, the frequency of the acoustic emission peak depends on the size of the objects, so using smaller absorbers might require the use of a different transducer, depending on the exact sensitivity bandwidth. Conversely, when increasing the bandwidth of the transducer, one should make sure that the optically absorbing objects are generating the right ultrasound frequency content. In the case of the uniformly absorbing plane, its thickness has to be adjusted to match the ultrasound frequency range of the transducer [21]. To ensure that the generated acoustic frequency content will be mostly driven by the size of the absorbing object, the stress confinement must be verified [22], and the laser pulse duration $\tau_{pulse}$ must therefore be short enough:

$$\tau_{pulse} < \tau_{ac} = \frac{D_a}{c_s}$$

where $D_a$ is the typical length of the absorber and $c_s$ the speed of sound. For a typical laser pulse duration of 10 ns, the stress confinement regime will be reached when the absorber diameter is much larger than 15 µm.

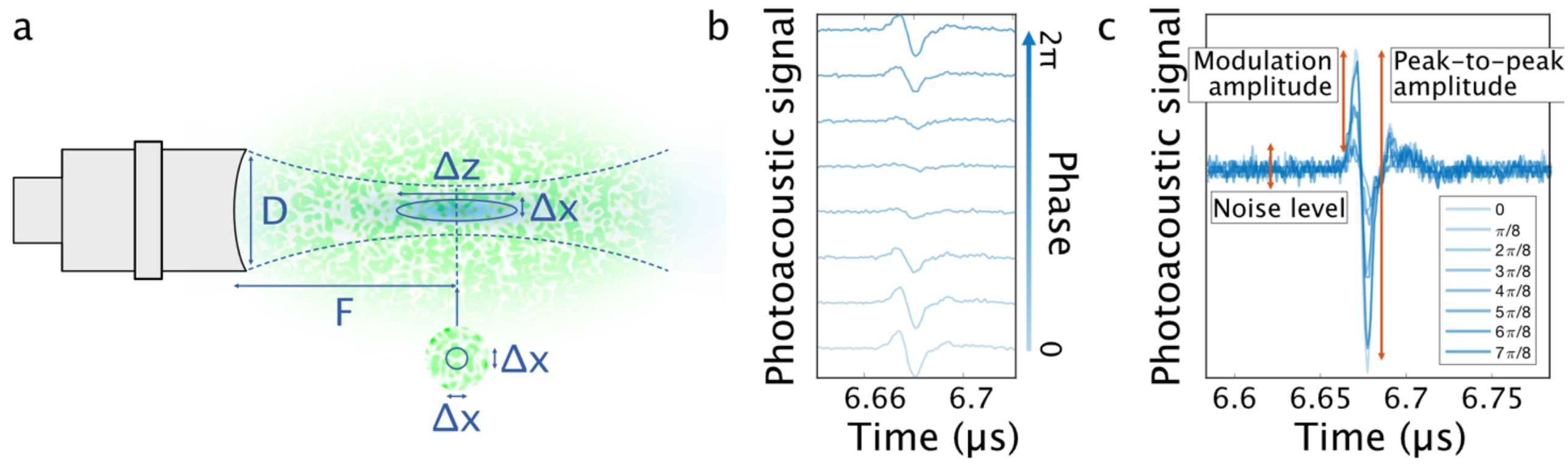

**Figure 3 - Key parameters influencing detection of PA signal modulation.** **a**, Typical specifications of a single-element focused ultrasound transducer: D diameter, F focal length, $\Delta x$, $\Delta z$: dimensions of the diffraction limited sensitivity region. Bottom: cross-section of the speckle pattern in the focal plane of the transducer to illustrate the mismatch between the dimensions of the sensitivity region and the speckle grains. **b**, Typical modulation of the PA feedback signal when modulating the phase of one input optical mode on the SLM. We consider here a transducer with a 10mm focal length. **c**, Illustration of the main metrics to assess.

### 3.3 Maximum enhancement

When considering a feedback signal linearly related to the PA signal, we need to consider the spatio-temporal constraint. Spatially, the maximum enhancement of the focused intensity varies as $\alpha \frac{N_{SLM}}{N_{speckle}}$, where $N_{SLM}$ is the number of SLM pixels and $\alpha = 0.5$ for a phase modulation SLM in a configuration with experimental noise [23,24]. Millions of pixels would therefore be ideal to obtain a significant enhancement *inside* scattering tissue. Temporally, due to the fast tissue decorrelation time (down to a few milliseconds) [25], we need to balance the trade-off between the enhancement and the duration of the focusing process (see 'How to choose suitable equipment' section).

Alternatively, fewer speckle grains can be probed by using sparse smaller absorbing objects and/or using higher ultrasound frequencies, as acoustic resolution scales as the inverse of these frequencies. In both cases, this would lower the number of probed speckle grains, but would lower the total optical absorption at the same time. This would effectively reduce the PA signal itself and thus the SNR. Detecting higher ultrasound frequencies with the same ultrasound transducer would also result in a lower PA signal amplitude due to a potential lower sensitivity. As ultrasound attenuation increases linearly with frequency in biological tissue [26,27] (approximately 1 dB per cm per MHz), these components would also be more subject to attenuation, further reducing the PA signal. This solution must therefore be considered with care, and the performances assessed depending on the exact experimental scenario. We note that these considerations apply in all wavefront shaping techniques where multiple speckle grains are probed.

## 4. How to choose suitable equipment

As discussed in the previous sections, the actual specifications of the hardware will drastically affect the performances of the focusing operation. To assist potential researchers interested exploring in the field, we provide some more insights into which features should be carefully considered. Supplementary table 1 details the various pieces of equipment (and their key specifications) that have been used so far in relevant works, including (1) excitation laser source, (2) spatial light modulator, (3) ultrasound sensor, (4) radio-frequency (RF) amplifier, (5) digitizer, (6) scattering media (e.g. diffuser). We also indicate in supplementary table 2 the focusing performances that have been achieved with these different setups. We believe this will help future researchers to build the optimal system given the constraints listed in the previous section.

### 4.1 Laser

One of the very first questions one encounters when designing an optical system for PA wavefront shaping pertains to the choice of a suitable laser source. On the one hand, the laser must serve as an adequate PA excitation source, while on the other it also needs to be compatible with the specific requirements for wavefront shaping as detailed below.

#### 4.1.1 Properties relevant for PA generation

*Pulse duration*

PA generation usually relies on pulsed excitation, employing pulse durations typically of 1 to 10 ns such that both thermal and stress confinement are observed.

*Repetition rate*

Pulse repetition rates are usually limited to maximum ~10 kHz to avoid interference between the acoustic signal and echoes from preceding pulses. In water, ultrasound will propagate over a distance of 15 cm during the inter-pulse period of 100 μs, which is larger than the usual targeted depths.

*Pulse energy*

Required pulse energies depend greatly on the class of PA imaging. In PA microscopy, the excitation is typically focused down to the optical diffraction limit, thereby reaching high optical fluences with pulse energies in the order of μJ. This technique is however limited to shallow depths in tissue, as any other laser scanning microscopy technique based on optical signals. At a target depths of several millimeters, scattered photons dominate over ballistic, and light propagates in the diffusive regime. Larger pulse energies of the order of mJ are therefore required, such that sufficient PA signal is generated under highly-scattered excitation (see Fig. 4a). The cost of laser sources grows quickly with pulse energy, especially when hand in hand with additional requirements on beam quality and coherence. It is thus highly valuable to provide more precise estimations on the minimum pulse energy requirement. We provide below a first order model to assess this energy constraints, as this has never been explicited to the best of our knowledge.

We consider that each laser pulse thermally induces a local and instantaneous increment in pressure $p_0$ in a small spherical volume (of diameter $d_0$) located at a depth $z$ inside the medium, as illustrated in Fig. 4a. This increment in pressure propagates as an attenuated spherical wave, reaching an ultrasound transducer at the surface of the sample with an amplitude $p_{UST} = p_0 d_0 \exp(-\alpha z)/(2z)$, where $\alpha$ is the acoustic attenuation coefficient. Note that the center frequency $f_c$ of the transducer sets the length scale $d_0$ as $0.66.c_s/f_c$, where $c_s$ is the speed of sound. Figure 4b shows the local pressure increase $p_0$, as a function of depth, that is required such that the pressure $p_{UST}$ at the detector matches its noise-equivalent pressure, considered as 0.5 $mPa.Hz^{-1/2}$ [28]. We assume a speed of sound $c_s$ = 1561 $m.s^{-1}$ and acoustic attenuation $\alpha$ = 0.54 $dB.cm^{-1}.MHz^{-1}$ typical of soft tissues [27]. Next, we estimate the optical fluence $\Phi_0$ required to induce such local pressure increments through $p_0(z) = \Gamma \mu_a \Phi_0(z)$, where $\mu_a$ denotes the optical absorption coefficient of the absorber, and $\Gamma$ the Grüneisen coefficient of its surroundings. To estimate the minimum pulse energy $E$ required at the sample surface, we assume that the radius of the excitation beam is given approximately by the depth [19], which propagates with effective optical attenuation coefficient $\mu_{\mathrm{eff}}$, such that $E = \Phi_0 exp(\mu_{eff} z) \pi z^2$. This is plotted versus depth in Fig. 4c, where we consider the absorber to consist of blood ($\Gamma$ = 0.14, $\mu_a$ = 22.48 $mm^{-1}$ at 532 nm, $\mu_a$ = 0.15 $mm^{-1}$ at 660 nm) at varying depths inside soft tissue (0.8103 $mm^{-1} \leq \mu_{\mathrm{eff}} \leq$ 1.416 $mm^{-1}$ at 532 nm, 0.0472 $mm^{-1} \leq \mu_{\mathrm{eff}} \leq$ 0.1089 $mm^{-1}$ at 660 nm).

*Wavelength*

The choice of emission wavelength may restrict, or be restricted by, the absorption band of the target absorber. Endogenous chromophores, such as hemoglobin, melanin or even water, can be either used as absorbing target or could alternatively perturb light penetration deep inside tissue [11]. To minimize endogenous absorption when using exogenous targets, optical wavelengths in the so-called optical window are typically used, between 600 nm and 800 nm.

PA imaging systems often employ tunable-wavelength sources (e.g. based on an optical parametric oscillator (OPO)), however the combination with high beam quality and coherence requirements makes the development of such laser sources a technically challenging task.

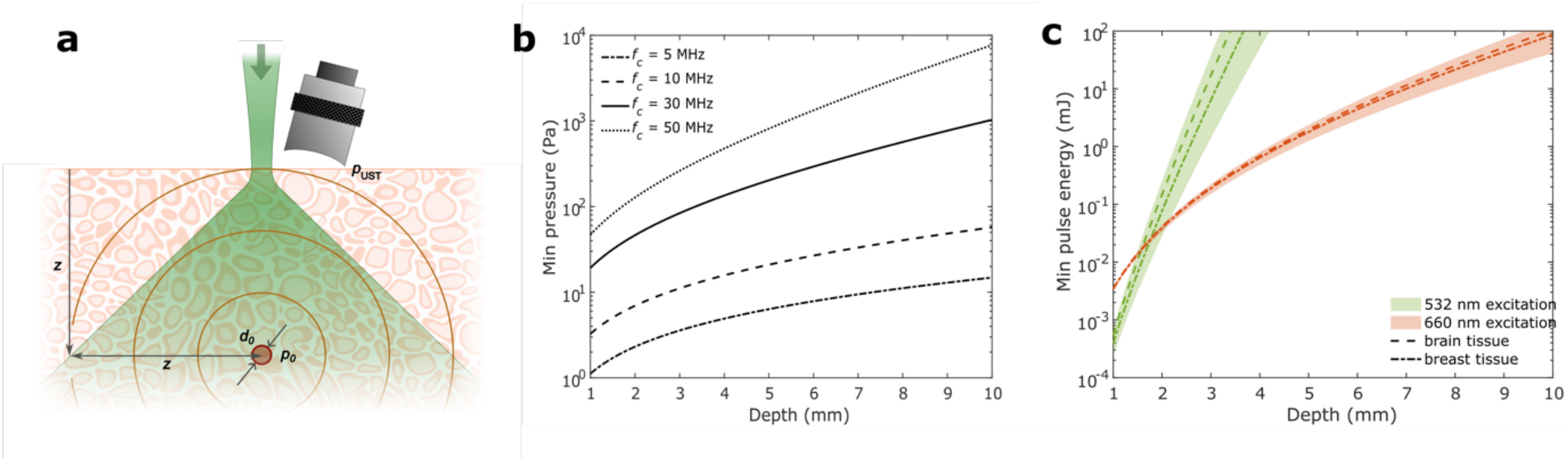

**Figure 4 – Modeling the required pulse energy for the laser source. a,** Schematic depiction of the model, where pulsed laser excitation induces an increase $p_0$ on the local pressure within a small volume of size $d_0$ at a depth $z$, which propagates as an acoustic spherical wave whose pressure $p_{UST}$ at the surface is measured by an ultrasound transducer. **b,** Required pressure increment $p_0$ such that $p_{UST}$ matches the noise-equivalent pressure of the detector, as function of depth and for transducers of varying center frequency $f_c$. **c,** Minimum pulse energy, as a function of depth, to generate the required pressure increment $p_0$ at $f_c$ = 30 MHz detector frequency, considering absorption (assuming 1% volume fraction) after diffuse propagation through soft tissues, at the 532 nm (green lines) and 660 nm (red lines) wavelengths. The dashed and dash-dotted lines denote scattering through brain and breast tissue (with 1% blood content), respectively.

### *4.1.2 Properties relevant for wavefront shaping*

*Mode stability*

Wavefront shaping aims to control the optical propagation behind (and ultimately *within*) complex media such as biological tissue by modulating the phase and/or the amplitude of a light beam – usually by means of a spatial light modulator. Since it fundamentally relies on such imprinted phase relations to be maintained throughout propagation across the complex medium, a high degree of coherence (spatial and temporal) is paramount. Regarding spatial coherence, laser sources emitting in only one transverse mode – usually the fundamental $TEM_{00}$ mode – with good beam quality ($M^2 \sim 1.5$) are widely available. This ensures in principle also adequate pulse-to-pulse repeatability and adequate speckle contrast. However pointing instabilities might arise with high power sources used in PA imaging, an issue that is usually not encountered in pure optical wavefront shaping experiments. These instabilities originate from thermal fluctuations in the laser cavity, and can drastically affect the performances of the shaping process, as this yields uncontrolled fluctuations in the speckle patterns, and therefore in the PA feedback signal.

*Coherence length*

The question of required temporal coherence is, however, a less obvious one [29]. To successfully apply wavefront shaping techniques in deep tissues or other complex media, the coherence length of the laser must meet certain requirements. This parameter is crucial because it ensures the laser light maintains its phase over the necessary distances, enabling accurate manipulation and focusing of light through scattering environments. While critically important for practical applications like deep tissue imaging and focusing, this aspect has been somewhat underexplored in existing research. In practical experiments, special attention needs to be paid to the laser's coherence length. The minimum coherence length that is required for wavefront shaping is determined by the distribution of path lengths travelled by photons in the complex medium, and as such depends on its scattering properties as well as the target focusing depth. We describe below a simplified model to estimate the coherence length requirements.

We assume isotropic diffusion from an infinitely short-pulsed point source located at a depth $z$ given by the transport mean free path $l_t' = 1/\mu_s'$ of a scattering medium with reduced scattering coefficient $\mu_s'$, as depicted in Fig. 5a. In this case, the diffusion equation has an analytical solution, with examples of the photon flux for several distances $r$ from the point source plotted in Fig. 5b. Note that the time integral of these curves decrease with $r$ due to accumulating optical absorption. The highlighted areas under the curves represent the fraction of photons arriving within the coherence time $\tau_c$ of the source (assumed 1 ns for illustration purposes). As one would expect, the fraction of photons arriving within the coherence time of the source – those which can be controlled coherently by wavefront shaping – decreases with increasing distance inside the scattering medium. Conversely, if one wishes the fraction of controlled photons to remain constant (e.g. at 90%) at increasing depths, the required temporal coherence for the source increases accordingly. In Fig. 5c, we show the estimated minimum coherence length $l_c = c.\tau_c$ (where c is the speed of light) required to ensure a fraction of controlled photons larger than 90%, for a typical range of reduced scattering coefficients found in biological media.

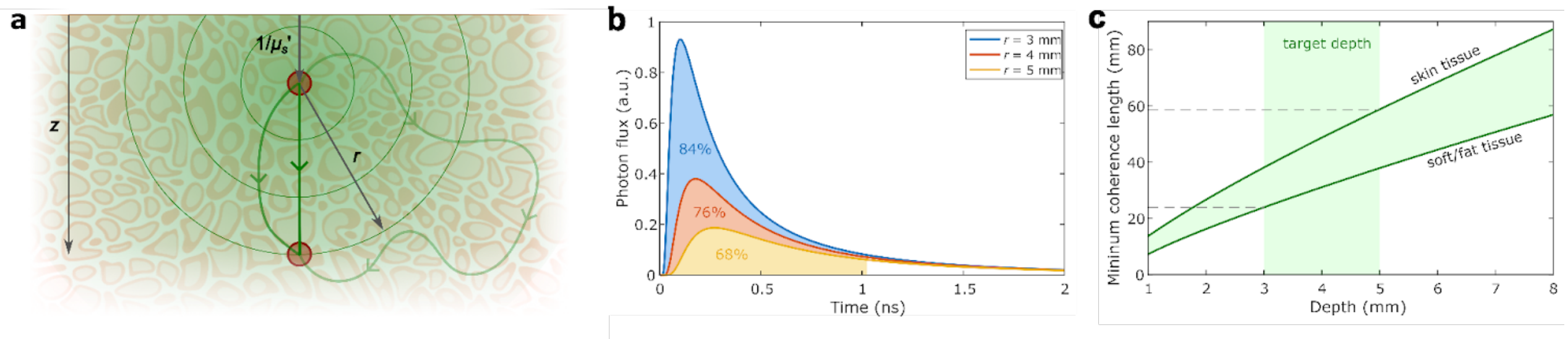

**Figure 5 – Modelling the required temporal coherence for the laser source. a,** Illustration of the model where optical propagation is approximated to isotropic photon diffusion from an instantaneous point source located at a depth $z = l_t'$ (one transport mean free path) inside biological tissue. **b,** Time-trace of the photon flux at fixed distances $r$ from the point source. The highlighted areas under the curves represent the fraction of photons arriving within a time interval of 1 ns. The parameters used are: $n$ = 1.4, $\mu_a$ = 0.002 mm$^{-1}$, and $\mu_s$ = 5 mm$^{-1}$. **c,** Minimal coherence length required to achieve control over $\eta$ = 90% light as function of depth, for a typical range of biological tissues. The parameters used are $n$ = 1.4, $\mu_a$ = 0.02 mm$^{-1}$, and $\mu_s$ from 1.75 mm$^{-1}$ (soft/fat tissue) to 4.21 mm$^{-1}$ (skin).

Nanosecond pulsed laser sources suitable both for PA generation and wavefront shaping applications in realistic scenarios still remain to be developed. The long coherence length of typical 532nm pump sources is usually considerably reduced after the OPO, which is needed to reach excitation wavelengths properly tuned to the optical absorption peak of the targeted chromophores. However, seed injection solutions exist and can restore coherence by narrowing down the optical linewidth of the emission [30].

**TIP box 1 - How to test the laser**

To confirm that the laser is suitable for wavefront shaping, insert a scattering medium (e.g. 120 grit ground glass diffuser) in the beam path, and place a camera behind it such that the transmitted speckle reaches the camera sensor (no problem if it is overfilled), at a distance such that the speckle grain (typically 10 to 100 µm) spans over several camera pixels. Take care to attenuate the laser beam appropriately (e.g. using ND filters or a pair of crossed polarizers) so as to not cause damage to the camera sensor. Use the camera in global shutter mode, set it to be triggered externally by the laser sync signal, set the exposure time to the time period between pulses (inverse repetition rate of the laser), and the trigger delay to half of this value – in this way, one obtains a speckle realization for individual laser pulses. Compute the cross-correlation between pairs of speckle realizations. This value should be as close to 1 as possible, indicating that pulses are highly repeatable in amplitude and phase distributions. Measure also the contrast of each speckle realization as the ratio between the standard deviation and average value of the intensity histogram. This ratio is indicative of its spatial coherence and should be as high as possible (usually >90% is desirable) as it corresponds to the maximum fraction of optical power which can be optimized by wavefront shaping. Low values for either of these two metrics generally arise from transverse multimode operation of the laser, and can in principle be improved by spatially filtering (focussing the beam through a pin hole of suitable diameter) to isolate the fundamental $TEM_{00}$ mode.

### *4.2 Spatial light modulator*

As discussed in the 'Experimental constraints' section, the focusing enhancement grows linearly with the amount of SLM pixels that can be independently addressed. However, the total duration of the wavefront correction grows linearly with this number as well. One must therefore find the right trade-off between pixel count and refresh rate to achieve a significant enhancement in a reasonable amount of time. This is to be considered in all optical wavefront shaping applications, but is even more critical when using PA feedback, as many speckle grains are probed.

This is of particular importance when attempting to focus inside scattering biological tissue, as speckle decorrelation can occur in under a second due to their dynamic nature, owing to intrinsic fluctuations sources at various scales, from intracellular motion to blood flow and breathing. We briefly describe here the existing technologies.

The most widespread technology in optical wavefront shaping is based on liquid crystal on silicon (LCoS) [31]. These devices combine a large pixel count in the order of 100k-1M, and can perform phase modulation with at least a 8-bit depth (256 levels). They however exhibit a low refresh rate of about 60 Hz, and modulate only one linear polarization. This can ultimately lead to a power loss, which can be tolerated in most wavefront shaping experiments, but would severely affect the amplitude of the PA feedback signal in our specific case.

As an example, a typical full HD SLM [32] contains 1920 x 1080 = 2 0736 00 pixels, which could yield a maximum enhancement of about 80 in the realistic scenario introduced before, but would require 1920 x 1080/60 = 34560 s > 9h to focus using the conventional optimization algorithm given the 60Hz refresh rate.

Employing SLM in off-axis configuration (i.e. isolating the first diffraction order) can in principle improve the signal-to-noise ratio by filtering out unmodulated light, yet at the cost of power [17].

Deformable mirrors can achieve a much faster phase modulation up to 25 kHz [33], at the cost of pixel count. The largest ones indeed only contain up to a few thousands, and therefore cannot be used to focus light when summing thousands of speckle grains within the focal region of the ultrasound transducer. Additionally, the cost of this solution can be prohibitive.

Digital micromirror devices (DMD) combine high speed (up to 20 kHz) and large pixel count of 0.1 to 1M pixels (see contribution by Popoff et al. in this special issue). They however are limited to binary amplitude modulation, which ultimately lead to a lower focusing enhancement [34]. Phase modulation can also be implemented with these devices [35], at the cost of pixel count and laser power, which again are of critical importance in our specific application. Additionally, the damage threshold of these devices is much lower than for LCoS SLM, up to two orders of magnitude [36]. This would again limit the available excitation power for PA generation.

High speed (up to 10 kHz) megapixel phase light modulators with 4-bit modulation depth have been recently developed [37]. This is achieved through a piston motion, as opposed to binary tilt motion in DMD. This device could be a good alternative to previous ones.

### *4.3 Ultrasound sensor and electronics*

Different types of ultrasound sensor can be used to detect the PA feedback signal. As described in detail in section 'Experimental constraints', the focusing performances are mostly limited by the number of speckle grains contained in a single resolution voxel and the capacity to resolve small amplitude modulations of the PA feedback signal. Regardless of sensor type, the first is essentially driven by its central frequency and bandwidth. The second depends on the noise-equivalent pressure level that characterizes the sensitivity of the detector [28], which can be associated with additional amplifiers. Filters can be introduced to also reduce the noise outside the useful frequency band. The bit depth of the digitizer should be appropriately chosen to provide sufficient dynamic range and resolve the feedback signal modulation, even in the absence of noise. For instance, a 12 bit digitizer can only encode the voltage signal from the ultrasound transducer on 4096 quantization levels: the modulation amplitude should therefore be larger than one level. Finally, cables are often overlooked but should be carefully chosen, to prevent noise pickup and reflections due to impedance mismatch. The working frequency range should also be adjusted to ensure best performances and increase the signal-to-noise ratio.

As detailed in supplementary table 1, only ultrasound transducers based on piezoelectric elements have been used so far in PA-guided wavefront shaping experiments. One could however benefit from the recent development of optical detectors of ultrasound [38–40]. These sensors exhibit a better sensitivity and larger bandwidth, potentially enabling to probe PA signal modulation from a reduced number of speckle grains. This arises from the miniaturization of the sensing elements, which does not cause substantial sensitivity reduction as with piezoelectric sensors [41]. Optical transparency of sensors based on Fabry-Perot cavities [42] also allow for simpler and more realistic geometry, as the excitation nanosecond pulsed laser can be transmitted through such a sensor.

**TIP box 2 - Good practice for sample mounting**

Acoustic coupling (or impedance matching) is a crucial aspect of ultrasound detection. This means that the ultrasound sensor should be mechanically coupled to the optically absorbing objects. The acoustic properties of the material should be comparable to water to ensure good propagation of the ultrasound waves. The agarose gel block holding the absorbing objects is usually immersed in water, along with the ultrasound transducer (as in Fig.2 d,e). Alternatively, ultrasound gel can be used when the experimental constraints prevent from immersing the various components in water (as in Fig.2 c). The presence of any bubble or hard surface in the vicinity of the absorbing objects should be prevented, which implies that the tank should be of larger size than the targeted depth. This will allow for clear time gating of the acoustic signals, thus removing potential reflections from walls.

In some geometries where the excitation laser beam will hit the surface of the acoustic sensor potentially causing damage (such as in Fig.2 c,d), a mylar film (i.e. survival blanket) can be used to reflect the laser beam while transmitting the acoustic waves.

### *4.4 Sample*

In PA wavefront shaping experiments, the proper preparation of samples is crucial for achieving effective imaging and focusing. Agarose gel is commonly employed as a mounting medium due to its versatility. Within this gel, various absorptive objects such as wires, beads, leaves, or ink either applied on slides or embedded directly can emulate experimental targets. These absorbers are key for generating detectable PA signals. Additionally, incorporating dyes or chromophores, which may be photoswitchable, into the gel enhances the PA contrast, aiding in the differentiation between the targets and the surrounding medium based on their optical properties. Contrasts with realistic scenarios where volumetric scattering and absorptive samples are spatially intermixed, in proof-of-principle experiments, it is common that a thin scattering layer (such as a diffuser) and an absorptive sample, separated by millimetres to centimetres, are utilized as a synthetic scattering media and sample combination. A summary of samples from representative previous work is provided in Table 3.

To prepare the agarose gel, a specific concentration of agarose powder is dissolved in water, typically between 1% and 2%, adjusted according to the desired optical scattering characteristics. This solution is heated until the agarose fully dissolves, then cooled to 50-60°C before embedding the absorptive objects or dyes. The mixture is then poured into molds and allowed to solidify, creating a stable, homogeneous medium suitable for PA experimentation. To mimic tissue scattering, intralipid solutions of varying concentrations can also be added to the gel [43].

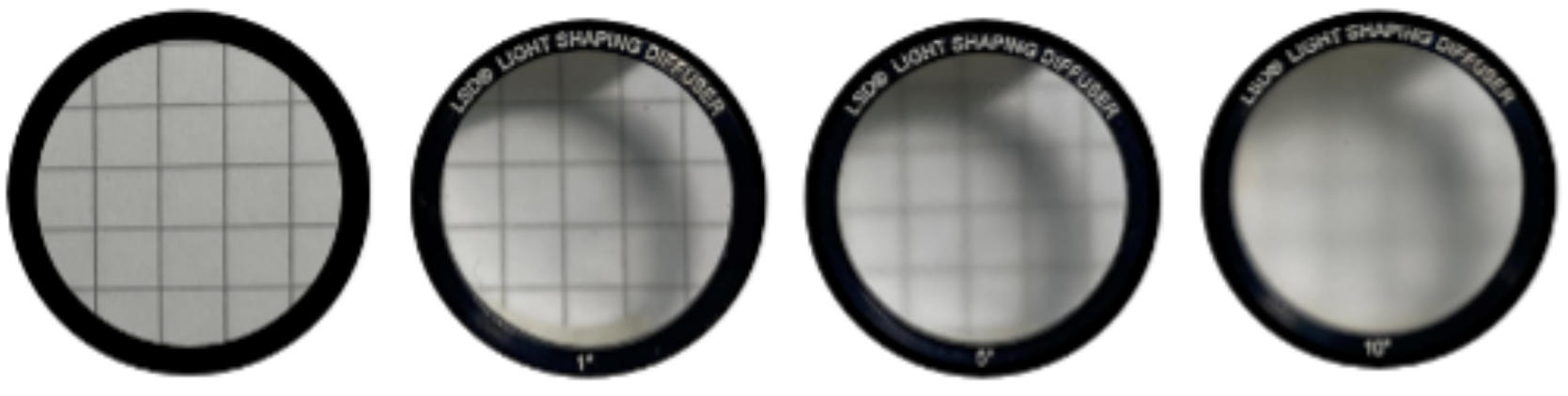

**Figure 6 – Holographic diffuser under different degrees showing different opaqueness** (1 cm away from the grid. Left to right: no diffuser, 1-degree diffuser, 5-degree diffuser and 10-degree diffuser)

## 5. Simulation

To help design the experimental setup and choose the appropriate parameters as well as equipment, we provide a Matlab script [44] to simulate light focusing in a scattering medium with PA feedback. The simulated configuration is the following:

- a flat-phase optical wavefront is impinging on a spatial light modulator (SLM) (phase modulation only).
- a camera is located the Fourier plane of the SLM and measures the light intensity of the produced speckle.
- an acoustic transducer is measuring the total light intensity within a region of interest, with a weighting profile mimicking the sensitivity profile.

We wrote the simulation so that researchers can directly choose the physical parameters of interest (optical wavelength of the light source, center frequency and aperture of the ultrasound transducer) and study how these would influence the outcome of the optimization process. Thanks to this numerical tool, we hope to make it easier to identify the right experimental configuration and their respective trade-offs to identify suitable configurations and get the best focusing performances.

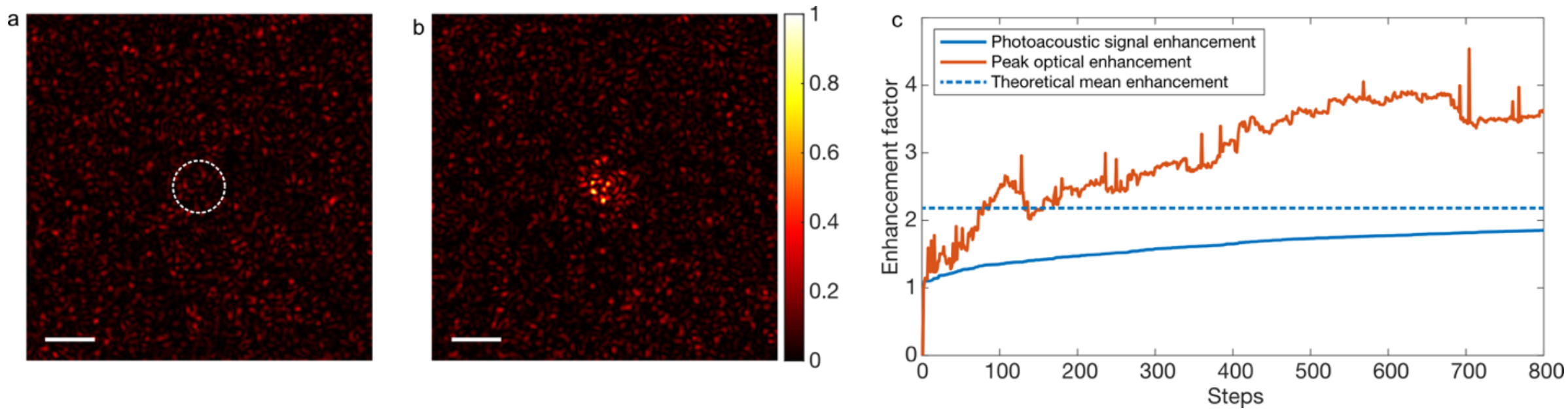

**Figure 7 - Numerical simulation of light focusing in a scattering medium by wavefront shaping with optimization of a PA feedback signal. a,** Initial speckle pattern before optimization. The dotted circle represents the -3dB threshold of the weight profile applied to the speckle pattern to compute the feedback signal. **b,** Speckle pattern after optimization. **c,** Enhancement: (solid blue) optimized PA feedback signal, (red) optical peak intensity within the optimized region, (dotted blue) theoretical mean enhancement of the PA feedback. Parameters were chosen as follows: optical wavelength: 1000 nm; SLM pixels: 400; central acoustic frequency: 100 MHz; numerical aperture of the transducer: 5. Scale bars: 5 µm

## 6. Which photoacoustic feedback signal?

In wavefront shaping experiment, metric design is crucial as it determines the underlying optimization problem. In the case of PA-guided wavefront shaping, we need to consider how to design such a metric as a feedback parameter from the recorded PA signals to utilize in the wavefront optimization procedure/algorithm as this turns out to be a critical component in the overall experimental scheme.

### *6.1 Pre-processing*

Prior to analyzing the PA signals, it is important to note that in general their measurements are strongly affected by detection noise, and that their magnitude is often affected by pulse-to-pulse variations in the excitation laser. The latter can be corrected by normalizing each measurement by the power of the corresponding excitation pulse, which can be measured independently using a beam splitter and fast photodiode (see Fig.2a). Detection noise consists mostly of uncorrelated white gaussian noise, i.e. random signal with constant power spectral density over a large bandwidth. The measurements may also contain additional parasitic signals radiated by high power equipment and picked up by the cables, or caused by electrostatic noise of the analog-to-digital converter and amplifier – all of which can interfere or be confused with the PA pulses (see for instance [36]). To improve their SNR, signals can be filtered prior to amplification and/or digitally after being recorded by an analog-to-digital converter to remove noise outside the spectral region of interest (dictated by the absorber dimensions and detection bandwidth). Another straightforward approach consists in averaging the signal over $N$ measurements – typically from few to tens of pulses – which increases the SNR by $\sqrt{N}$ in the presence of white noise. Despite being a very effective way to mitigate noise, averaging quickly increases the time required for the wavefront optimization procedure owing to such square root dependence. Thus, strategies to improve SNR requiring fewer measurements have been introduced, such as digital processing based on wavelet denoising and correlation detection [36,45,46].

To avoid hindering the acquisition speed, fully analog signal processing schemes can provide a viable alternative, albeit at the cost of increased hardware complexity. A sophisticated PA interrogation scheme based on lock-in detection was demonstrated to improve the SNR by an order of magnitude [47]. It should be noted that this scheme requires both modulation (acousto-optic modulator) and de-modulation (boxcar integrator, lock-in amplifier) hardware.

### *6.2 Linear feedback*

Upon pre-processing (filtering, averaging, denoising) the time-trace of a PA signal from an isolated absorber typically resembles that shown in Fig.8a. The exact shape of the signal depends on several factors, from the characteristics of the object (dimensions, geometry, or more generally its volumetric distribution of absorption coefficient), to the spatial distribution of the (speckled) illumination pattern itself, and characteristics of the detection (sensitivity profile, and spectral bandwidth). The most straightforward feedback signal is using the peak-to-peak amplitude of the PA signal. As a first approximation, this value is proportional to the optical intensity impinging on the light-absorbing object emitting the pressure wave – restricted to the sensitivity region of the ultrasound detector.

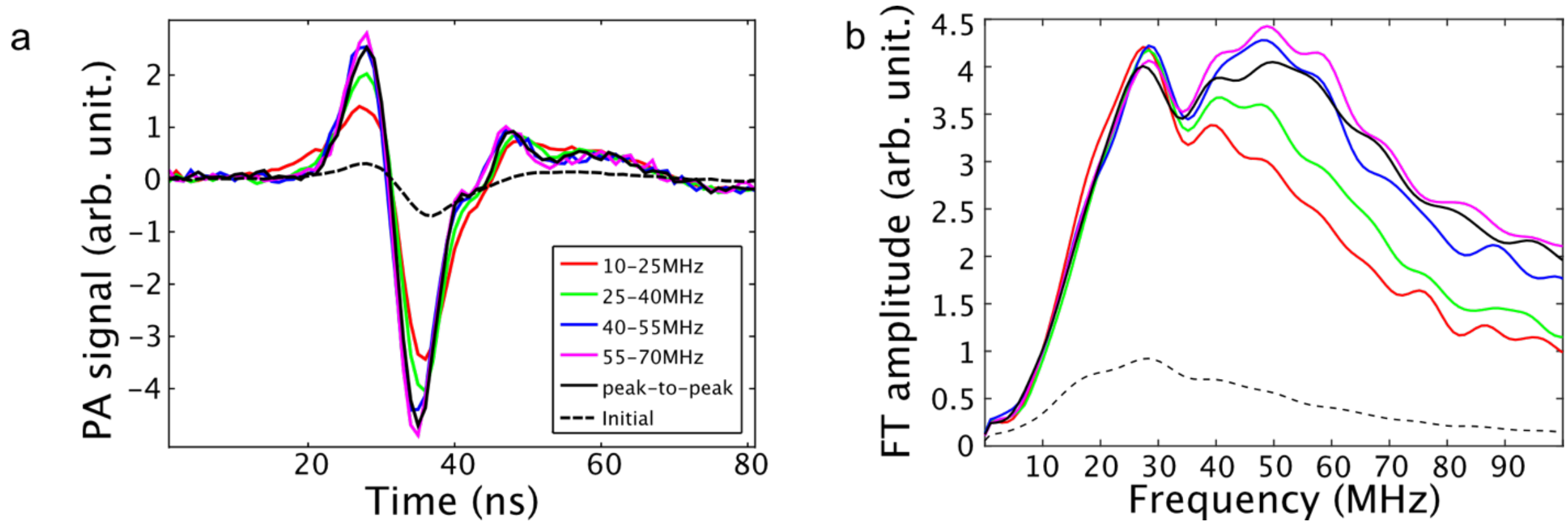

**Figure 8 - Photoacoustic feedback signal. a,** Photoacoustic signal prior to (dashed line) and after (solid line) optimizing its peak-to-peak value with optical wavefront shaping. **b,** Effect of using the signal within different frequency bands (highlighted in color) as the feedback in the wavefront shaping optimization procedure. The absorber consisted in a 30µm-diameter black nylon wire, illuminated with speckle with 25µm grain size, and detected using a 27MHz spherically focussed ultrasound transducer.

Although the spectrum carries the same information as the signal, the frequency-domain representation can however make some of it more easily accessible. The size of the acoustic focal volume is inversely proportional to the ultrasonic frequency and can therefore be controlled by appropriate spectral filtering. To this end, the RMS value of the spectrum over a selected bandwidth can be conveniently employed as the feedback signal [14]. Figure 8b shows the spectra of the optimized signals, resulting from optimizations on four different spectral bandwidths, comparing them with the average initial spectrum (i.e. of the unoptimized signal) as well as the spectrum resulting from choosing the peak-to-peak amplitude as the feedback mechanism. Optimization of the high-frequency components of the PA signals was shown to increase the enhancement and to result in tighter optical focusing, since the number of speckle grains decreases as the probed acoustic volume is reduced.

### *6.3 Nonlinear feedback*

In the linear regime, ultrasound detectors sense the integrated pressure within their focal volume, with little sensitivity to how pressure (and hence light intensity) is distributed within it. As a result, PA-guided wavefront optimization based on linear feedback signals generally results in coarse light confinement at the scale of the acoustic diffraction limit. In other words, the focused illumination contains multiple speckle grains (see Fig.7b). Nonlinear feedback mechanisms can in principle favor the accumulation of excitation light into smaller volumes, thus offering a promising route to further improve focusing down to the optical diffraction limit, i.e. confining the laser excitation to a single speckle grain.

A nonlinear relation between the light intensity and the amplitude of the induced PA signal typically manifests under high optical peak power, and depends also on the optically absorbing sample [48]. Different sources of such PA nonlinearities include saturation and damage, thermal nonlinearities, and phase transformations [47]. The lock-in detection approach mentioned above is particularly suited for extracting nonlinear signals, by demodulating the signal at harmonics of modulation frequency.

Interestingly, it can be possible to find or generate nonlinear feedback mechanisms even in the absence of material nonlinearities. One such strategy exploits the temperature dependence of the Grüneisen parameter of an absorbing sample upon dual-pulse excitation [16]. Here, a first pulse generates an acoustic signal, while also increasing the temperature locally. Thus, a second pulse arriving within the thermal confinement time (~200 μs) encounters an increased Grüneisen parameter, and as such gives rise to an acoustic signal stronger than the first. Lai et al. showed that the difference in amplitude between the two pulses increases nonlinearly with optical fluence, and used it as feedback mechanism to focus light to a single speckle grain on an extended absorber, reaching an optical enhancement of 6000 – a record still holding nearly a decade later.

Another approach takes advantage of the varying absorbing distribution from samples consisting of flowing absorbers. The inherent fluctuation of their PA response allows optimizing the standard deviation of the signals, which increases nonlinearly as the illumination volume decreases [49]. This resulted in sub-acoustic focusing down to a single speckle grain, and a 4-fold increase in optical enhancement with respect to optimizations employing PA amplitude feedback. It should be noted that the previous two approaches make use of a nonlinear feedback, despite the actual PA signals being generated linearly with optical fluence.

It is also worth noting that sub-acoustic focusing does not always require a nonlinear feedback mechanism. The spatial response of focused ultrasound transducers is non-uniform, giving higher weighting to acoustic signals originating closer to their center. Under certain experimental conditions – employing a very high frequency transducer, high-pass filtering the signals at 80MHz, and with few speckle grains within its acoustic focus – this was shown to result in light focusing beyond the acoustic diffraction limit [15], even though the PA amplitude was used as the feedback signal.

### *6.4 PA-image feedback*

Above we have considered that the acoustic interrogation comprised a single detector (an ultrasound transducer), and discussed how to extract a feedback parameter from the PA signal generated by a single pulse, or several consecutive pulses. An array of detectors [50,51] opens the possibility of reconstructing PA tomographic images through a multitude of algorithms, such as back-projection or back-propagation of the time-domain signals [52]. In this framework, one can look for a suitable feedback mechanism in the PA images rather than in time-domain signals. A two-dimensional detector array allows reconstructing a volumetric PA image for every excitation pulse. From each such three-dimensional image, one can employ as a feedback mechanism the maximum intensity within a chosen volume-of-interest to guide an iterative wavefront optimization which seeks to maximize light delivery into that region [53].

A different approach made use of reconstructed PA images to measure the PA transmission matrix [10,13]. As for the optimization procedure described before, varying the phase of each successive SLM input pattern causes the intensity values of the corresponding PA image to vary according to the cosine of the applied phase. This allows to determine the phase and amplitude of the elements of each column of the PA transmission matrix. This method allows parallelizing the wavefront optimization over all PA image voxels, offering selective light focusing at any target location on the absorbing object [13]. This method can actually also be used with single-element focused ultrasound transducers [10].

## 7. Discussion - Challenges to focus light *inside* tissue

In this tutorial, we have provided a detailed description of how to perform a basic proof-of-principle experiment in PA-guided wavefront shaping. To the best of our knowledge, all related works have only performed light focusing *through* scattering layers.

As we move forward, the transition from proof-of-principle experiments to practical focusing *within* biological tissues emerges as a significant interest area, particularly in biomedical applications where deeper focusing is desired. However, several challenges remain to be addressed. A primary issue is the substantial mismatch between optical and acoustic spatial resolutions, which complicates the extraction of faint modulations of the PA signal within voxels filled with numerous optical speckle grains. Despite a decade of research efforts exploring this avenue, focusing deep inside tissue has not yet been achieved. To address this challenge, a precise estimation of the required signal-to-noise ratio (SNR) and dynamic range is crucial. These parameters are fundamental for accurately detecting and interpreting the fluctuations of PA signals induced by optical wavefront modulations amid the inherent background in tissue environments. For a detailed understanding and numerical analysis of these requirements, we discuss SNR and dynamic range estimation in Section 'Experimental constraints'.

Furthermore, it is essential to acknowledge that advancements in hardware are crucial for improving PA-guided wavefront shaping. It mostly depends on the actual speed of the wavefront shaping itself but also on how to detect very weak PA signals and modulations. To clarify the needs for hardware developers, we outline several desirable features for the hardware involved in such experiments. This includes 1) a fast spatial light modulator (SLM) capable of operating at frequencies greater than one kilohertz and with megapixel resolution. Moreover, 2) laser sources that operate in the near-infrared (NIR) spectrum, ideally tunable, with kilohertz repetition rates and short (< 2 ns) pulse durations, with a few centimeters coherence length, are important for achieving the high ultrasound frequencies necessary for deep tissue imaging. These lasers must also provide sufficient energy per pulse to ensure robust signal generation while complying with safety standards that limit laser intensity to under 200mW/cm^2 at the tissue (which usually translates to about 1 W direct output power). 3) The development of high-frequency ultrasound detectors with high sensitivity are equally critical, requiring high signal-to-noise ratio for effective signal detection. The use of optical ultrasound sensors could be of particular interest, as it can provide better sensitivity at ultra high frequencies (> 50 MHz) compared to conventional piezoelectric sensor

Together, these technological advancements are integral to developing an advanced PA wavefront shaping system capable of focusing light within tissue.

Building on the challenges of achieving precise focusing within tissues, it's crucial to explore what can still be accomplished when direct focusing is not feasible. Recognizing that focusing is not the sole pathway to obtaining meaningful images opens up alternative strategies for enhancing PA imaging capabilities. One promising approach is to guide light in a less precise, but effective manner towards deeper structures within tissues. By estimating the potential increase in light intensity as a function of depth, it is possible to improve light delivery to targeted areas, thereby enhancing the overall imaging process. Additionally, the concept of focusing light for therapeutic purposes, such as using targeted light to ablate tissue, presents a valuable application of PA wavefront shaping. This method leverages the precise control over light to perform minimally invasive procedures, offering a synergistic blend of imaging and therapeutic capabilities. These alternative strategies not only extend the utility of PA imaging beyond traditional focusing but also underscore the versatility of light manipulation in biomedical applications, promising significant advances in both diagnostics and treatment methodologies.

## Acknowledgements

FX, IL and RP acknowledge the Chan Zuckerberg Initiative (Deep Tissue Imaging grants no. 2020-225346 and 2024-337799).

We thank Nikita Kaydanov and Samuel Davis for helpful discussions and feedback.

## Table 1: Equipment list of partial published PA wavefront shaping works

| Ref. | Equipment | | | | |
|---|---|---|---|---|---|
| | **Source** | **SLM** | **Detector** | **Amplifier** | **DAQ** |
| **[9]** | custom Cr,Nd:YAG 532nm, 1ns, 100Hz | DM Boston Micromachines | 75MHz focussed UST 40MHz focussed UST | 60dB | oscilloscope |
| **[15]** | Nd:YAG frequency doubled Continuum Surelight I20 532nm, 5ns, 20Hz, 21uJ (after SLM) | LCoS-SLM BNS 512x512 | 90MHz focussed UST (Olympus V3512, 50MHz bandwidth) | 40dB (low-noise) Femto HAS-X-2-40 | oscilloscope |
| **[10]** | Continuum Surelight 532nm, 5ns, 10Hz, 10mJ | DM (Boston Micromachines Multi-DM) 12x12, 2kHz | 28MHz focussed UST (Sonaxis SNX110509_HFM13) | Sofranel 5900PR | oscilloscope (Lecroy WaveSurfer 104 MXs-B) |
| **[13]** | Continuum Surelight 532nm, 5ns, 10Hz, 10mJ | DM (Boston Micromachines Multi-DM) 12x12, 2kHz | linear US array (128 elements) 14.4MHz, -6dB bandwidth 6.8MHz (Vermon) | | commercial US scanner Supersonic Image, Aixplorer |
| **[14]** | Continuum Surelight 532nm, 5ns, 10Hz, 10mJ | DM (Boston Micromachines Multi-DM) 12x12, 2kHz | 27MHz focussed UST (Sonaxis, -6dB pulse-echo bandwidth 27MHz) | Sofranel 5900PR | oscilloscope (Lecroy WaveSurfer 104 MXs-B) |
| **[54]** | Nd:YLF EdgeWave BX-series 532nm, 1kHz, 800uJ | DMD TI D4100, 1024x768 pix | 10MHz foccused UST (Panametrics A315D, -6dB bandwidth 5.5MHz) | 100x (Minicircuits ZFL-500LN) | oscilloscope (Tektronix DPO2024) AlazarTech ATS9350 500Ms/s |
| **[16]** | EdgeWave Innoslab BX2II-E 532nm, 10ns, 0-30kHz, 0.2mJ | LCoS-SLM Holoeye Pluto | 50MHz focussed UST (non-focusing Panametrics NDT V358 + focussing lens) | 50dB (Mini-Circuits ZFL-500LN+ & ZX60-43-S+) | oscilloscope (Tektronix TDS5034) |
| **[53]** | Q-switched Nd:YAG Spectra Physics Lab-190-30 532nm, 6ns, 15Hz, 4mJ | LCoS-SLM Holoeye Pluto-BB II | spherical array 256 USTs 4 MHz, -6dB 100% | N/A | N/A |
| **[50]** | Q-switched Nd:YAG Spectra Physics Lab-190-30 532nm, 6ns, 15Hz, 4mJ | LCoS-SLM Holoeye Pluto-BB II | spherical array 256 USTs 4 MHz, -6dB 100%, 200um isotropic resolution | N/A | N/A |
| **[15]** | Nd:YAG frequency doubled Continuum Surelight I20 532nm, 5ns, 20Hz, 9.4uJ (after SLM) | LCoS-SLM BNS 512x512 | 90MHz focussed UST (Olympus V3512, 50MHz bandwidth) | | |
| **[47]** | Spectra-Physics Mosaic 532-11 532nm, 10ns, 20kHz, 150uJ | LCoS-SLM Meadowlark P512-532 | 20MHz focussed UST (Olympus V316, 15MHz 6dB) | | boxcar integrator (SRS SR250) |
| **[49]** | Q-switched diode-pumped Spectra-Physics Mosaic 532-11 532nm, 7ns, 20kHz | LCoS-SLM Meadowlark P512-532 | 20MHz focussed UST (Olympus V317) | N/A | high-speed digitizer |
| **[36]** | LABest SGR-10 532nm, 10Hz, 800uJ | DMD (TI DLP6500) | 5MHz UST (SIUI 5Z10SJ30DJ) | Minicircuits ZFL-500LN+ | oscilloscope (Lecroy WaveMaster 806Zi-A) |
| **[46]** | LABest SGR-10 532nm, 10Hz, 800uJ | DMD (TI DLP6500) | 20 MHzUST (Olympus V317) | Minicircuits ZFL-500LN+ | oscilloscope (Lecroy WaveMaster 806Zi-A) |
| **[45]** | LABest SGR-10 532nm, 10Hz, 800uJ | DMD (TI DLP6500) | 5MHz UST (SIUI 5Z10SJ30DJ) | Minicircuits ZFL-500LN+ | oscilloscope (Lecroy WaveMaster 806Zi-A) |
| **[55]** | LABest SGR-10 532nm, 5Hz, 800uJ | DMD (TI DLP6500) | 5MHz UST (SIUI 5Z10SJ30DJ) | Minicircuits ZFL-500LN+ | oscilloscope (Lecroy WaveMaster 806Zi-A) |
| **[56]** | Elforlight SPOT-10-200-532 532nm, 2ns | DMD (TI DLP700, 47kHz) | 50 MHz UST (Olympus V358) | Spectrum Instrumentation SPA.1411 | Spectrum Instrumentation M4i.4420 |
| **[57]** | Yenista Tunics T100S-HP (cw) AOM & EDFA 1440-1640nm, 75ns, 10kHz, 1.8uJ | LCoS-SLM Meadowlark HSP1920-600-1300 | 5 MHz focussed UST (Panametrics NDT V310-N-SU) | DAQ system w/ pulser-receiver (Olympus 5072PR) | |

## Table 2: Enhancement and relevant parameters in partial published PA wavefront shaping works

| Ref. | Metric used | Optimisation method | Speckle size (FWHM, µm) | UST focus size (FWHM, µm) | Number of modes | Number of SLM pixels used | $\eta$ Theoretical enhancement | $\eta$ Actual enhancement |
|---|---|---|---|---|---|---|---|---|
| **[9]** | peak-to-peak | iterative | 10 | 41<br>90 | 17<br>81 | 140 | 6.5<br>1.4 | 5 - 10 |
| **[15]** | peak-to-peak (after high-pass filtering) | genetic [15] (iterative) | 36 | 36 | 1 | 804 | | 10 |
| **[10]** | peak-to-peak (within 90ns time windows), 2-60MHz bandpass filter | TM phase-shifting recording in Hadamard basis | 25 | 100 | 6 | 140 | 11.5 | 6 |
| **[13]** | from 2D PA image intensity<br>low-passed 25MHz | TM phase-shifting recording in Hadamard basis | 25 | | 25 | 140 | 3 | 3 - 6 |
| **[14]** | integral in varying spectral bandwidths | iterative in Hadamard basis | 25 | 110 | 6.7 | 140 | 10 | 4 - 12 |
| **[54]** | peak-to-peak | Hadamard inversion binary amplitude | 550 | 400 | 1 | 1024 | 163 | 14 |
| **[16]** | peak-to-peak linear<br>peak-2-peak diff NL | genetic [15] (iterative) | 5 | 65 | 169<br>1 | 20736 | 97 linear | 60 linear<br>6000 NL |
| **[53]** | maximum of 3D PAT image within ROI (obtained at each pulse) | genetic [15]) (iterative) | 90 | 200 | 5 | 400 | 40 | 6 |
| **[50]** | maximum of 3D PAT image within ROI (obtained at each pulse) | genetic [15] (iterative) | 27 | 200 | 13.7<br>55<br>55 | 400 | 14.6<br>3.6<br>3.6 | 4.5<br>3.5<br>3.5 |
| **[15]** | peak-to-peak (after high-pass filtering) | genetic [15] (iterative) | 13 | 38 | 8.5 | N/A | N/A | 8.5 |
| **[47]** | lock-in detection | genetic [15] (iterative) | | 300 | 10 | 256 | 20 | 9 linear<br>16 NL |
| **[49]** | (i) peak-to-peak<br>(ii) stdev (randomly fluctuating absorbers) | genetic [15] (iterative) | 15.5<br>10.3 | 208 | 180<br>408 | 1024<br>4096 | | |
| **[36]** | peak-to-peak wavelet denoising + correlation detection | genetic [15] (iterative, binary amplitude) | 150 | 880 | 7.5 | 576 | 12 | 7.8 |
| **[46]** | peak-to-peak denoising with dynamic window | genetic [15] (iterative, binary amplitude) | N/A | N/A | N/A | N/A | N/A | 8 |
| **[45]** | peak-to-peak denoised | genetic [15] (iterative, phase+amplitude super-pixel) | 150 | 880 | 7.5 | 312 | 42 | 10.5 |
| **[55]** | peak-to-peak denoised | TM (phase+amplitude, super-pixel) | 150 | 880 | 7.5 | 312 | 42 | 10.75 |
| **[56]** | | RVITM real-valued intensity TM binary amplitude | | | 45 | 4096 | 14.5 | 7 |
| **[57]** | peak-to-peak | genetic [15] (iterative) | | | | | | 8 |

## Table 3: Sample list in partial published PA wavefront shaping works

| Ref. | Sample | | |
|---|---|---|---|
| | Diffuser | Absorber | distance (mm) |
| [9] | 2x layers paraffin 0.25mm | graphite (typewritter carbon tape)<br>10um/50um graphite particles | 2 |
| [15] | ground glass 120grit (Edmund) | India ink inside polypropylene tube | 8 |
| [10] | 0.5deg light shaping diffuser (Newport)<br>500um-thick chicken breast (partially dried, sandwiched between glass slides) | 30um black nylon wires (VetSuture NYL02DS 10/0) in agarose gel | |
| [13] | ground glass (Thorlabs DG10-120) | black leaf skeleton embedded in agarose gel | |
| [14] | 0.5deg light shaping diffuser (Newport) | 30um black nylon wires (VetSuture NYL02DS 10/0) in agarose gel | 60 |
| [54] | ground glass (Thorlabs DG10-120) | black ink in silicone tube (Silastic, 1mm diameter) in water<br>500um-microspheres (Phosphorex 1500KR, red polystyrene) | 140 |
| [16] | ground glass (Thorlabs DG10-120) | whole blood | 10 |
| [53] | ground glass (Thorlabs DG10-120) | 200-um black paramagnetic PE microspheres (Cospheric BKPMS 180210) in clear agar | 35 |
| [50] | ground glass (Thorlabs DG10-120) | 100um and 200um PE microspheres (Cospheric BKPMS 90-106 and BKPMS 180-210), 400um carbon microspheres (SPI-Supplies) in clear agar | 55 |
| [15] | NA | 25um & 50um black alpaca hair | |
| [47] | NA | black tape | 20 |
| [49] | RPC Photonics EDC-1 | 3um red-dyed polystyrene microspheres (SigmaAldrich 42922) in tube | |
| [36] | ground glass 120 grit (Edmund 83419) | 150um black nylon thread in agarose gel | 30 |
| [46] | ground glass 120 grit (Edmund 83419) | 100um black nylon thread in agarose gel | 25 |
| [45] | ground glass 120 grit (Edmund 83419) | 150um black nylon thread in agarose gel | 16 |
| [55] | ground glass<br>(Daheng Optics GCL-201103) | 150um black nylon thread in agarose gel | 10 |
| [56] | ground glass 220 grit<br>(Thorlabs) | black tape | 5 |
| [57] | ground glass (Thorlabs DG10-120) | absorbing layer | |